\documentclass[twocolumn,preprintnumbers,amsmath,amssymb]{revtex4-1}
\usepackage{graphicx}
\usepackage{dcolumn}
\usepackage{bm}
\usepackage{color}

\begin{document}

\title{Shocks near Jamming}

\author{Leopoldo R. G\'omez$^{1}$}
\email{gomez@lorentz.leidenuniv.nl}
\author{Ari M. Turner$^{2}$}
\author{Martin van Hecke$^{3}$}
\author{Vincenzo Vitelli$^{1}$}
\email{vitelli@lorentz.leidenuniv.nl}

\affiliation{ $^1$Instituut-Lorentz
for Theoretical Physics, Leiden University, Leiden NL 2333 CA, The Netherlands. \\
$^2$Department of Physics, University of California, Berkeley,
California 94720, USA.\\
$^3$Kamerlingh Onnes Lab, Universiteit Leiden, Postbus 9504, 2300
RA Leiden, The Netherlands.}

\date{\today}

\begin{abstract}
Non-linear sound is an extreme phenomenon typically observed in
solids after violent explosions. But granular media are different.
Right when they jam, these fragile and disordered solids exhibit a
vanishing rigidity and sound speed, so that even tiny mechanical
perturbations form supersonic shocks. Here, we perform simulations
in which two-dimensional jammed granular packings are dynamically
compressed, and demonstrate that the elementary excitations are
strongly non-linear shocks, rather than ordinary phonons. We
capture the full dependence of the shock speed on pressure and
impact intensity by a surprisingly simple analytical model.
\end{abstract}

\maketitle

Granular materials
exhibit a wide range of complex collective behaviors, making them
an important testing ground for the physics of amorphous materials
\cite{Epitome}-\cite{Xu2010}. The confining pressure $P$ is
perhaps the most important parameter controlling their properties.
Strongly compressed granular media are, in many aspects, simple
solids in which perturbations travel as ordinary phonons. However,
when the confining pressure is lowered to zero, or the amplitude
of the disturbance is much higher than the initial
compression, the mechanical response of granular media becomes
increasingly anomalous.

Several insights have been obtained by studying a simple model of
granular media comprised of soft frictionless spheres just above
the jamming point \cite{Epitome}-\cite{Xu2010}. The jamming point
corresponds to the critical density at which the grains barely
touch and $P$ vanishes \cite{Epitome}. The first insight is that
the vibrational modes of jammed packings resemble ordinary phonons
only below a characteristic frequency scale $\omega^*$ that
vanishes as $P$ goes to zero \cite{Wyart}-\cite{Souslov}. Above
$\omega^*$, the modes are extended but strongly scattered by
disorder \cite{Xu}-\cite{mattransport}. Second, as a direct
consequence of the nonlinear dependence of the local contact force
on the grain deformations, the sound speed vanishes as $P$ goes to
zero \cite{Nagel}-\cite{mattransport}: linear sound cannot
propagate when the particles barely touch. Third, the range of
validity of linear response vanishes when $P$ goes to zero. This
is intuitive since the material is about to fall apart when the
pressure vanishes \cite{Xu2010}.

As the pressure (or density) is lowered towards the jamming point, there are thus three anomalies that
forbid the propagation of ordinary phonons: disorder disrupt phononic transport for all
frequency scales, the sound speed vanishes and linear response is
no longer valid. The vanishing of the sound speed and absence of a
linear range clearly suggest that the excitations near jamming
will be strongly nonlinear. Nevertheless, most numerical and
analytical studies of energy transport have been carried out in solids just above the jamming point, within a
vanishingly small window of linear response. By explicit design, these studies cannot probe non-linear energy transport because the
dynamics of the system is solved through a normal mode expansion
\cite{Xu,mattransport,Vitelli,Somfai}. Therefore, with the exception of
theoretical and experimental studies on solitons in one
dimensional granular chains, started with the seminal work of
Nesterenko \cite{NesterenkoBook,Nesterenko1984,Sen,
Daraio,DaraioII}, non-linear energy transport in granular packings
remains largely unexplored.

\noindent {\it Numerical model.} \quad \!\!\!\!  To probe how
elastic energy is transported close to the jamming point, we
performed molecular dynamics simulations of a piston-compression
experiment carried out in two dimensional polydisperse amorphous
packings of soft frictionless spheres, whose radii, ${R_i}$, are
uniformly distributed between 0.8 and 1.2 times their average $R$.
Particles  $i$ and $j$ at positions $\vec{x}_i$ and $\vec{x}_j$ interact
via a non-linear repulsive contact potential \cite{Somfai}:
\begin{equation}
V_{ij}=\frac{\varepsilon_{ij}}{\alpha}\,\,\delta_{ij}\,^{\alpha}
\label{eq:potential}
\end{equation}
only for positive overlap $\delta_{ij}\equiv  R_i + R_j-|\vec{x}_i
-\vec{x}_j|>0$, otherwise  $V_{ij}=0$ when $\delta_{ij}\le0$. Here, the
interaction parameter $\varepsilon_{ij}=\frac{4}{3}\frac{R_i
R_j}{R_i + R_j} E^{*}_{ij}$ is expressed in terms of the effective
Young's modulus of the two particles, $E^{*}_{ij}$, see Ref.
\cite{Somfai} for more details. The case $\alpha=5/2$ corresponds
to Hertz's law. Lengths are measured in units of average particle
diameter. The unit of mass is set by fixing the grain density to
unity. The effective particle Young modulus $E^{*}$ is set to one,
which becomes the pressure unit. These choices ensure that the
speed of sound \emph{inside} the grain, $v_g$ is one
\cite{Somfai}.

\begin{center}
\begin{figure}[t]
\includegraphics[width=8.5 cm]{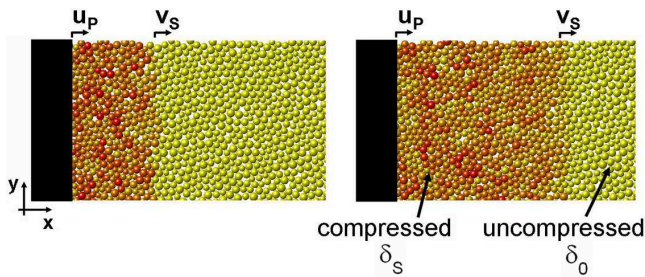} \caption{
Snapshots of the piston-compression simulation. A massive piston
moves to the right at a constant velocity $u_P$, resulting in the
formation of a compression front traveling at a speed $v_S$. Color
indicates the local pressure at each grain. The average particle
overlap is $\delta_0$ to the right of the front and
$\delta_S>\delta_0$ to the left.} \label{schematic}
\end{figure}
\end{center}

We prepare Hertzian packings at a fixed pressure $P$, or
equivalently, an average particle overlap $\delta_{0} \sim
P^{2/3}$. They are then continuously compressed by a piston which
moves with a constant velocity $u_P$ in the $x$ direction
throughout the simulation, see Fig. ~\ref{schematic}. The
subsequent motion of the particles is obtained by numerical
integration of Newton's equations of motion subject to periodic
boundary conditions in the $y$ direction and a fixed boundary on
the right edge of the system. We use two dimensional packings in
the range of $10^3$ to $10^4$ particles with various width to
length ratios.

\begin{center}
\begin{figure}[b]
\includegraphics[width=8.5 cm]{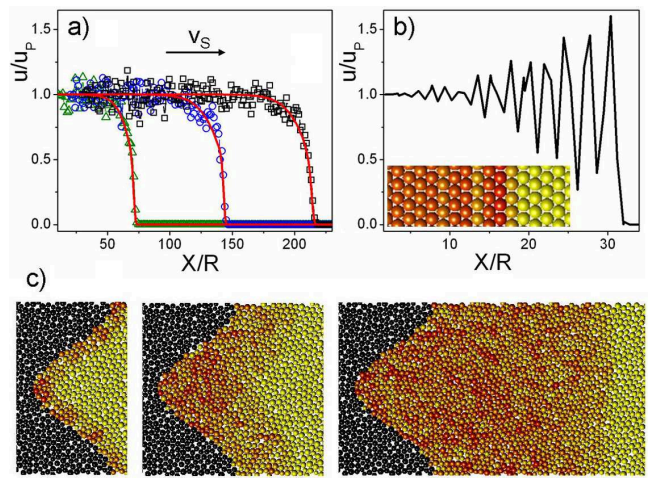} \caption{
(a) Profiles of a shock wave at different times obtained by
averaging the particle velocity in the $\hat{y}$ direction
(symbols). The red lines shows the fits of the fronts to an
empirical fit formula. (b) Oscillatory velocity profile of a
shock-like wave propagating through an hexagonal array. The inset
shows the in-phase motion of
the crystalline planes that gives rise to the pressure oscillations. (c) Representative snapshots of the
focusing and flattening of an initially curved front generated by
a sinusoidal piston (time progresses from left to right).}
\label{schematic2}
\end{figure}
\end{center}

\noindent {\it Phenomenology.}  \quad \!\!\!\! The piston
compression leads to the formation of a front that separates two
states. Ahead of the front, we find a region where the particles
are at rest having the initial overlap $\delta_0$, whereas behind
it there is a compressed region with particles moving on average
with the piston speed $u_P$ and an overlap $\delta_S > \delta_0$.
Figure ~\ref{schematic2}a shows typical profiles for the
\emph{longitudinal} particle velocity $u$ (in the $\hat{x}$
direction) as a function of $x$, obtained upon averaging velocity
fluctuations in the $\hat{y}$ direction.

Two qualitative features of the shocks stand out for all the
amorphous packings probed in this study: the fronts are smooth and
stable. The smoothness can be contrasted with the typical shock
profile that arise in ordered lattices of grains. Figure
~\ref{schematic2}b, obtained for a triangular lattice of grains
with zero initial overlap, shows large coherent pressure
oscillations caused by the in-phase motion of the crystalline
planes. These peaks are washed out by disorder in the amorphous
packings.

Second, we have systematically tested the stability of the front
against sinusoidal perturbations (in the $y$-direction) of varying
amplitudes and wavelengths in disordered packings under various
pressures. This was done through directs simulations \cite{Gomez}
as well as by performing a Dyakov's stability analysis
\cite{Gomez,Dyakov,Swan}. A typical result from our simulations,
illustrated  in Fig.~\ref{schematic2}c, shows how the front
remains stable due to a classic stress focusing process, where
particles ``left-behind'' experience a large compression, pushing
them to catch up with the rest of the front. In the light of these
observations, the shocks can be treated as one dimensional front
propagation phenomena.

\begin{figure}
\includegraphics[width=8.5 cm]{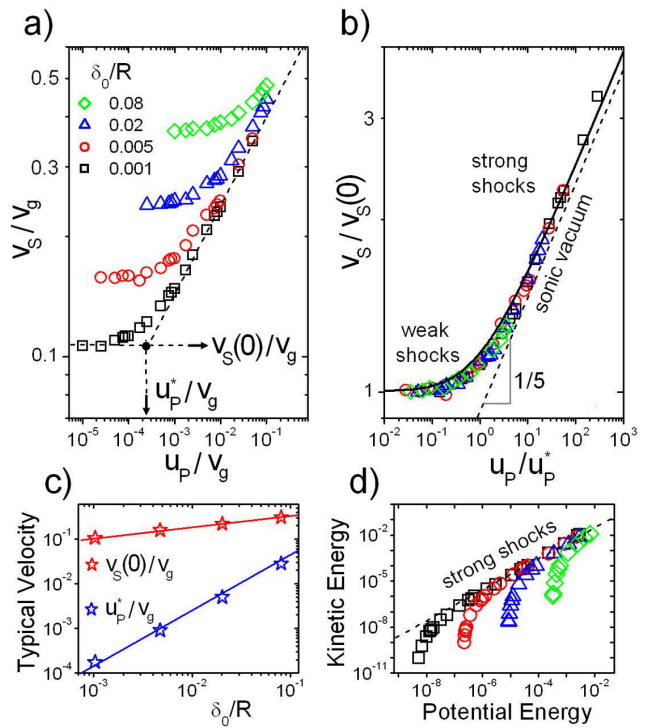} \caption{
a) Speed of the front $v_S$ versus particle velocity $u_P$
measured in units of $v_g$, the sound speed within the grain, for
decreasing particle overlap $\delta_0$. b) Same plot as in (a) but
with $v_S$ normalized by $v_S(0)$ and $u_P$ normalized by the
crossover particle speed $u_P^\star$: $v_S(0)$ and $u_P^\star$ are
indicated in panel (a). The dashed line indicates the power law
$v_S \sim {u_P}^{1/5}$ characteristic of a sonic vacuum. The black
line indicates the theory developed here to describe the universal
transition from weakly to strongly non-linear waves in systems
close to jamming. c) Variation of $v_S(0)$ and $u_P^*$ with
distance to the jamming transition parameterized by the initial
average overlap $\delta_0$. The dashed lines indicate the power
laws $v_S(0) \sim \delta_0^{1/4}$, $u_P^\star \sim
\delta_0^{5/4}$. d) Variation of the kinetic energy with potential
energy in dimensionless units - same color code as in (a-b). The
dashed line indicates the linear relationship observed for strong
shocks.} \label{hugoniot}
\end{figure}

\noindent {\it Front speed.} \quad \!\!\!\!  Once transients have
died out, the front propagates with constant speed $v_S$ in the
amorphous packings. Upon using conservation of mass across the
shock front,  we derive a one dimensional relation between the
characteristic velocities $u_P$ and $v_S$, through the average
radius of the particles, $R$, and the average compression in the
shock $\delta_S$, and ahead of it, $\delta_0$:
\begin{equation}
v_S= u_P\, \frac{2R-\delta_0}{\delta_S-\delta_0} ~. \label{tt}
\end{equation}
Since the particle compression $\delta_S$ is typically much less
than its diameter $2R$, Eq. (\ref{tt}) implies that $v_S \gg u_P$.
This is consistent with our numerical findings summarized in
Fig.~\ref{hugoniot}a where the dependence of $v_S$ on $u_P$ is
explored systematically for different compressions.

Inspection of Fig.~\ref{hugoniot}a reveals two distinct regimes.
For low $u_P$, the front speed $v_S$ is nearly independent of
$u_P$ - in this (quasi)linear regime, $v_S$ is simply controlled
by the initial pressure $P$. The strongly non-linear shock waves
regime is reached for high compression speed $u_P$, where $v_S$
depends on $u_P$, but not on $P$.

The data for $v_S$ can be collapsed onto a single master curve, as
shown in Fig.~\ref{hugoniot}b. We achieve this upon rescaling the
$v_S$ axis by $v_S(0)$, the numerically determined value that the
front speed attains in the limit of vanishing $u_{P}$ (see
Fig.~\ref{hugoniot}a). The $u_P$ axis is rescaled by a
pressure-dependent velocity scale $u_{P}^{*}$, obtained by
matching the low and high $u_P$ asymptotes in Fig.~\ref{hugoniot}a
(see arrow): $u_{P}^{*}$ marks the crossover between linear
acoustic waves and shocks.

\noindent {\it Scaling analysis.} \quad \!\!\!\!  The pressure
dependence of $v_S(0)$ can be rationalized using scaling
arguments. We expect that $v_S(0)$ reduces to $c$, the speed of
linear longitudinal sound waves. To determine the scaling of $c$
with pressure, note that $c \sim \sqrt{B}$, where the bulk modulus
$B=\frac{\partial P}{\partial V}$ and $P=\frac{\partial
E}{\partial V}$. The change in volume $dV$ scales linearly with
$\delta_0$, the average overlap between particles, while the
energy scales as $E \sim \delta_{0}^{\alpha}$, see Eq.
\ref{eq:potential}. Upon setting $\alpha=5/2$, we obtain the
pressure dependence of the longitudinal speed of sound
$c\sim\delta^{1/4}_{0}  \sim  P^{1/6}$ valid for Hertzian
interactions \cite{Somfai}. Figure \ref{hugoniot}c shows that the
numerical data for $v_S(0)$, represented by red symbols, is
consistent with the $\delta_0^{1/4}$ scaling, which is shown as a
continuous red line.

We now turn to the regime of high piston speeds, $u_P \gg
u_P^*$, when the front speed $v_S$ becomes nearly independent of $P$. Since
$u_P$, $R$ and $\delta_0$ are all known, we need one additional
relation which combined with Eq.~(\ref{tt}) will make a definite
prediction for the shock speed. We note that for strong shocks, the
propagating front generates a characteristic compression $\delta
\gg \delta_{0}$ and a corresponding increase in the kinetic
energy. By assuming that the kinetic and potential energies are of
the same order, we obtain $u_{P}^2 \sim \delta^{5/2}$. We have
tested numerically that this non-trivial proportionality relation
exists for strong deformations, see Fig. \ref{hugoniot}d.
Upon combining the balance between kinetic and potential energy with
Eq.~(\ref{tt}), one readily obtains the power law $v_S \sim u_P^{1/5}$, plotted
as a dashed line in Fig. \ref{hugoniot}b. This scaling relation is
clearly consistent with our numerical data for the speed of
strongly non-linear shock waves.

We deduce the dependence on compression of the crossover speed
$u_{P}^*$ by smoothly matching the two asymptotic relations for
the front speed $v_S \sim u_P ^{{1}/{5}}$ and $v_S(0) \sim
\delta_0 ^{{1}/{4}}$. This leads to the power law relation
$u_{P}^* \sim \delta_0^{5/4}$ (continuous blue line in Fig.
\ref{hugoniot}c) that is consistent with our numerical values
(blue symbols). Note that the data collapse in Fig.
\ref{hugoniot}b depends only on the scaling $u_{P}^* \sim
\delta_0^{5/4}$ and it is not sensitive on the precise definition
of the crossover speed. Upon using the conversion relation $u_{P}
\sim \delta^{5/4}$, the intuitive expectation that the crossover
takes place when $\delta \approx \delta_0$ is confirmed.

We conclude that by controlling $\delta_0$ or P, which
parameterize the distance to the jamming point (at $P=0$ and
$\delta_0=0$), we can tune $u_P^*$ and the onset of the strongly
non-linear response of the packings. Our key numerical findings on
the shock velocity summarized in Fig. \ref{hugoniot} can be
grasped from scaling near the jamming point.

\noindent {\it Analytical model.} \quad \!\!\!\! In order to
account for the dependence of $v_S$ on $u_P$ and the smoothness of
the shock profiles, we construct the simplest possible 1D model
that quantitatively accounts for the trends observed in Fig.
\ref{hugoniot} and sheds light on the role of disorder.

In the continuum limit, we obtain the following equation governing
the dynamics of the system in terms of the strain field
$\delta(x,t)$ \cite{comment}:
\begin{equation}
\frac{R^2}{3}\delta_{ttxx}-\delta_{tt}+\frac{4 R^2 \varepsilon}{m} \,
[\delta^{\alpha-1}]_{xx}=0 . \label{fg}
\end{equation}

To gain some intuition for the physics behind Eq. (\ref{fg}), note
that by setting $\alpha=2$, one recovers a linear dispersive wave
equation, with speed proportional to $\sqrt{\varepsilon/m}$ in the long wavelength limit. By
contrast, when $\alpha>2$ a non-linear wave equation is obtained.
Nonlinearities and dispersive effects gives
rise to finite amplitude waves: either solitary waves or shocks
are possible depending on the drive \cite{NesterenkoBook}.

\begin{figure}[tb]
\includegraphics[width=7.0 cm]{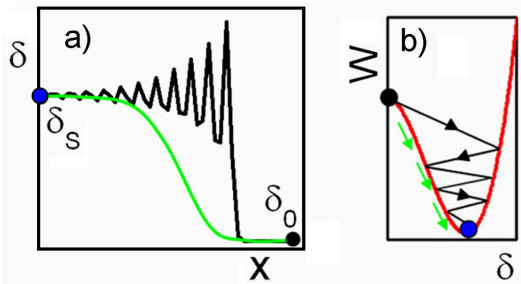}
\caption{(a) Simulations of an ordered chain with small viscosity
display large coherent oscillations in the front profile (black
line) \cite{Herbold}. If the viscosity is large enough, one obtains an
homogeneous shock profile, shown as a green line, similar to the profile in Fig. \ref{schematic2}a. (b) The presence
of an {\it effective} viscosity will induce the oscillation of the
particle (black trajectory) towards the bottom of the potential
$W(\delta)$, shown as a red line. If the viscosity is large
enough, the particle can move directly to the minimum without
performing any oscillations, see the green trajectory corresponding
to the homogeneous shock profile of panel (a). } \label{fig4}
\end{figure}

Shock propagation is modeled by the combined strain $\delta(x,t)=\delta_0 + g(\tilde{x})$, where $g(\tilde{x})$ gives the
shape of the shock and $\tilde{x} \equiv x-v_S\,t$. Upon inserting this
ansatz into Eq. (\ref{fg}), we obtain the conservation law
$\frac{1}{2} \delta_{\tilde{x}}^{2}+W(\delta)=0$, where
$W(\delta)$ is given by
\begin{eqnarray}
W(\delta)&=&\frac{24  \varepsilon }{m\alpha
v_S^2}\,(\delta^\alpha-\delta_0^\alpha)-\frac{3}{R^2}(\delta^2-\delta_0^2)\nonumber\\
&&-24\delta_0(\frac{\varepsilon}{m
v_S^2}\delta_0^{\alpha-2}-\frac{1}{4R^2})(\delta-\delta_0) .
\end{eqnarray}
This conservation law can be interpreted as describing the total
energy of an effective particle at position $\delta$ rolling down
a potential well $W(\delta)$, shown as a red line in Fig.
\ref{fig4}a (here $\tilde{x}$ maps to time so that $\frac{1}{2}
\delta_{\tilde{x}}^{2}$ is the kinetic term of the particle)
\cite{Herbold}.

One of the key ideas of our work is that disorder can act as an
effective viscosity for the shock: the energy imparted
unidirectionally by the piston is redistributed among other
degrees of freedom, reducing the energy propagating with the shock
front. In our mapping, this implies that the effective particle,
initially located at the maximum of the potential $W=0$, moves to
the minimum of the potential well (see Fig. \ref{fig4}a). Thus,
upon setting $\partial_{\delta}W(\delta)=0$, we can obtain a
relation between propagation velocity and induced compression in
the front
\begin{equation}
\frac{v_S}{c}=\sqrt{ \frac{1}{\alpha-1}\,
\frac{(\delta_S/\delta_0)^{\alpha-1}-1}{(\delta_S/\delta_0)-1}}
\label{th}
\end{equation}
that is independent of viscosity, even if an infinitesimal amount
of dissipation is necessary to obtain a steady state solution of
Eq. (\ref{fg}).

Together, Eqs. (\ref{tt}) and (\ref{th}) can be seen as a parametric
relation between front and particles velocities, where the overlap
$\delta_S$ produced by the passage of the front is the parameter.
Such a parametric plot of $v_S$ versus $u_P$ is drawn as a
continuous curve on the numerical data in Fig. \ref{hugoniot}b.
This comparison shows that Eqs.~(\ref{tt}) and (\ref{th}) are in
excellent agreement (without any fitting parameter) with the
results of our numerical experiments on shock propagation.

\noindent {\it Discussion.} \quad \!\!\!\! The shock formation
explored in the present study is a generic phenomenon independent
of the dimensionality of the sample that relies purely on the
presence of a nonlinear law between grains (for any $\alpha>2$)
and not on the presence of friction. Experimentally, this can be
tested by impacting a box of (frictional) glass beads with a heavy
mass, for a range of impact speeds and pressures - preliminary
experimental results for the front speed compare favorably to our
theoretical predictions in Fig. \ref{hugoniot} \cite{Martin}.

We note, however, that in frictional granular media, a second type
of densification front can be observed, which is often referred to
as plowing \cite{Jaeger,Mandi}. Whereas our shock waves always propagate
with speeds above the linear sound speed, and continue to
propagate even after the driving stops, plowing fronts are
generally much slower (in \cite{Jaeger}, of order 1 m/s), and stop
almost immediately when the driving stops. We believe that the
underlying difference is that our shocks are dynamical phenomena,
set by a  balance of potential and kinetic energies, whereas
plowing is in essence a quasistatic phenomenon, dominated by
dissipation. In the dynamic case, the change in packing fraction
induced by the shock is associated with grain deformations,
whereas in the quasistatic case, densification is dominated by
grain rearrangements and compaction.

{\it Outlook.} \quad \!\!\!\!  The shocks that arise in grains
near jamming are just one representative of a broader class of
strongly nonlinear excitations that emerge near the marginal state
of suspensions, emulsions, wet foams and weakly connected fiber
networks \cite{MvHecke}, \cite{mattPRL}, \cite{chase}. Close to
losing their rigidity, all these materials exhibit a vanishing
range of linear response, so that almost any amount of finite
driving will elicit an extreme mechanical response in the form of
rearrangements, yielding and flow \cite{Xu2010}, \cite{brianPRL},
\cite{Gollub}, \cite{Zaccone}. It remains an open question whether
all these phenomena can be successfully described in terms of
simple scaling near jamming.

\textbf{Acknowledgments} We acknowledge inspiring discussion with X. Campman, V. Nesterenko,
W. van Saarloos, A. Tichler and N. Upadhyaya. LRG and AMT acknowledge financial
support from FOM and Shell.


\begin{thebibliography}{9}
\bibitem{comment} Equation (\ref{fg}) is simpler than the one originally introduced
by Nesterenko  \cite{NesterenkoBook}.
\end{thebibliography}
\end{document}